\documentstyle[12pt,psfig]{article}
\newcommand{\be}{\begin{equation}}
\newcommand{\ee}{\end{equation}}

\newcommand{\bea}{\begin{eqnarray}}
\newcommand{\eea}{\end{eqnarray}}

\begin{document}
\title{``Soft'' transverse expansion and flow in a multi-fluid model
without phase transition}
\author{
{\bf J. Brachmann, A. Dumitru, M. Bleicher,}\\
{\bf J.A. Maruhn, H. St\"ocker, W. Greiner}
\\[0.3cm]
{\small Institut f\"ur Theoretische Physik der J.W.Goethe Universit\"at}\\
{\small Postfach 111932, D-60054 Frankfurt a.M., Germany}
\\[1cm]
{\small Proceedings for the}\\
{\small XXXV International Winter Meeting on Nuclear Physics}\\
{\small Bormio (Italy), February 3rd - 8th, 1997}\\
{\small (ed.\ :\ I. Iori)}
\\[2cm]
}
\maketitle   
\begin{abstract}
We study transverse expansion and directed flow in Au(11AGeV)Au reactions
within a multi-fluid dynamical model.
Although we do not employ an equation of state (EoS) with a first order 
phase transition, 
we find a slow increase of the transverse velocities of the nucleons
with time. A similar behaviour can be observed for the directed nucleon flow.
This is due to non-equilibrium effects which also lead to less and slower
conversion of longitudinal into transverse momentum.  
We also show that the proton rapidity distribution at
CERN energies, as calculated within this model, agrees well with 
the preliminary NA44-data.
\end{abstract}
\newpage     

\section{Introduction}
The success of the hydro model at BEVALAC energies 
(e.g.\ the prediction of flow) and its simplicity
encouraged to extend the 
hydrodynamical (one-fluid) model also to higher impact energies.
At very high baryon densities and / or temperatures a phase transition
from ordinary hadronic matter to a QGP is expected~\cite{StoePR,QGPtrans}. 
In the (one-fluid) hydrodynamical model the
energy and baryon densities necessary for this phase transition are already
reached in the BNL-AGS energy regime~\cite{Yaris}. Within this model,
the effects of a (first order-like) phase transition lead to
\begin{enumerate}
\item a local minimum in the excitation function of the collective
nucleon flow $\langle p_x^{dir}/N\rangle$~\cite{Yaris,otherflow},
\item a prolonged lifetime of the system at the ``softest point'' of the
EoS~\cite{Hung} due to a slower transverse expansion.
\end{enumerate}

The reason for this is that the EoS is softened in the phase coexistence
region as compared to an EoS without phase transition.

\section{The three-fluid model}
Of course, the question whether the phase transition region is reached
relies considerably on how much of the incident energy is deposited
in the reaction zone and converted to compressional and thermal energy. 
Due to the assumption of instantaneous local 
thermalization of projectile and target, it is clear that in the
one-fluid model the maximum possible energy is deposited
at midrapidity during the compressional stage.
On the other hand, it is very questionable that the assumption of
instantaneous thermalization holds in the ultrarelativistic
energy range. Considering the forward-backward peaking of the differential
$pp$ cross-section \cite{refppdNdy} the protons are shifted about one unit
in rapidity towards midrapidity (in $24~$GeV $pp$ collisions)
and thus can be treated as separated in rapidity even after the interaction. 
The same holds for the produced particles
-- mainly pions -- which are produced at midrapidity.

\subsection{Coupling source terms of the nucleonic fluids}
This motivates to build a hydrodynamical model with three different fluids
in order to account for the non-equilibrium between projectile, target
and produced particles in the early stage of a heavy encounter.
These fluids $1$, $2$, $3$ correspond to projectile and target nucleons 
and produced particles (called the {\em fireball}), respectively.
The individual energy-momentum tensors $T_l^{\mu\nu}$ and 
baryon currents $j_l^\mu$ do not need to be conserved,
since the three fluids may in principle exchange energy, momentum 
and baryon charge:
\begin{eqnarray}\label{HEq2b}
\partial_\mu T_l^{\mu\nu} = F_l^\nu \quad,
\quad \partial_\mu j_l^\mu = S_l \quad\quad (l=1..3)\quad.
\end{eqnarray}
Since the source terms $F_l^\nu$ denote energy/momentum loss of fluid
$l$ per volume and unit time, they can be parametrized by the collision rate
times energy/momentum loss in a single $NN$ collision \cite{2fluid}.
The source terms $S_l$ denote the creation or loss of baryons within fluid $l$.
As a consequence of the forward-backward peaking of the $pp$ cross-section
we neglect the baryon-exchange within the nucleonic fluids, $S_1=S_2=0$.
Since only the sum of the source terms $S_l$ and $F_l^\nu$ needs to equal zero,
the fireball also remains net baryon free, $S_3=0$, and the fireball
source term $F_3^\nu$ is obtained by $F_3^\nu=-F_1^\nu-F_2^\nu$.

In general it is always possible to split the source terms in a
symmetric and an antisymmetric part with respect to the fluid indices
($1\leftrightarrow2$):
\bea
\label{sourcesprinciple}
\partial_\mu T_1^{\mu\nu} & = & f_{exchange}^\nu -f_{loss}^\nu\quad, \nonumber\\
\partial_\mu T_2^{\mu\nu} & = & -f_{exchange}^\nu -f_{loss}^\nu\quad, \\
\partial_\mu T_3^{\mu\nu} & = & 2f_{loss}^\nu\quad.\nonumber
\eea
The antisymmetric term $f_{exchange}^\nu$ describes the exchange of energy and
momentum between projectile and target fluid, while $f_{loss}^\nu$ denotes the
loss of energy and momentum transferred to the fireball.

We compute these two terms like in the two-fluid model of
\cite{2fluid} from a parametrization of the mean
energy respectively longitudinal momentum loss in a single 
nucleon-nucleon collision.
By setting $f_{loss}=0$, it is possible to switch to a two-fluid model
without creating a fireball.
%

For a further reading on the three-fluid model we refer the reader
to \cite{3fluid,Dum95a}.

\subsection{One-fluid transition}
In the later stage of the collision the nucleonic fluids stop.
Their relative velocity is then comparable to the internal thermal velocities.
The two fluids are no longer separated in phase space, so that the main
assumption for a two-fluid region does not hold anymore.
Moreover, the coupling source terms cease to be valid,
since they do not account for thermal smearing and vanish linearly with
the relative velocity.
Since we do not account for thermal smearing, the two fluids are merged
into one, if the relative velocity is comparable to the
root-mean-square velocity in a nonrelativistic degenerate Fermi gas
or a nonrelativistic Boltzmann gas.
Presently, only the one-fluid transition 
of the {\em nucleonic} fluids is implemented.
%

\subsection{The EoS for the baryonic fluids}
The baryonic fluids are treated as a non-relativistic ideal gas 
with compression energy.
\be\label{EoS}
p = \frac{2}{3} (\epsilon-E_c n) +p_c \quad.
\ee
For the compression energy, we employ the ansatz
\be\label{comprene}
E_c=\frac{k_c}{18n n_0} (n-n_0)^2 +m_N +W_0 \quad,
n_0 \approx0.16~fm^{-3}\quad.
\ee
so that the compressional pressure $p_c$ is:
\be
p_c = -\frac{dE_c}{dn^{-1}}=n^2\frac{dE_c}{dn}=\frac{k_c}{18n_0}
(n^2-n_0^2) \quad.
\ee
We emphasize that neither a phase transition, nor heavy resonances,
nor attractive baryon-baryon interactions are included,
which would lead to a ``softening'' of the EoS in the one-fluid limit.
This will become essential when we compare the non-equilibrium
effects to the one-fluid limit.

\section{Non-equilibrium Effects}
\begin{figure}[htp]
\vspace*{-2cm}
\centerline{\hbox{\psfig{figure=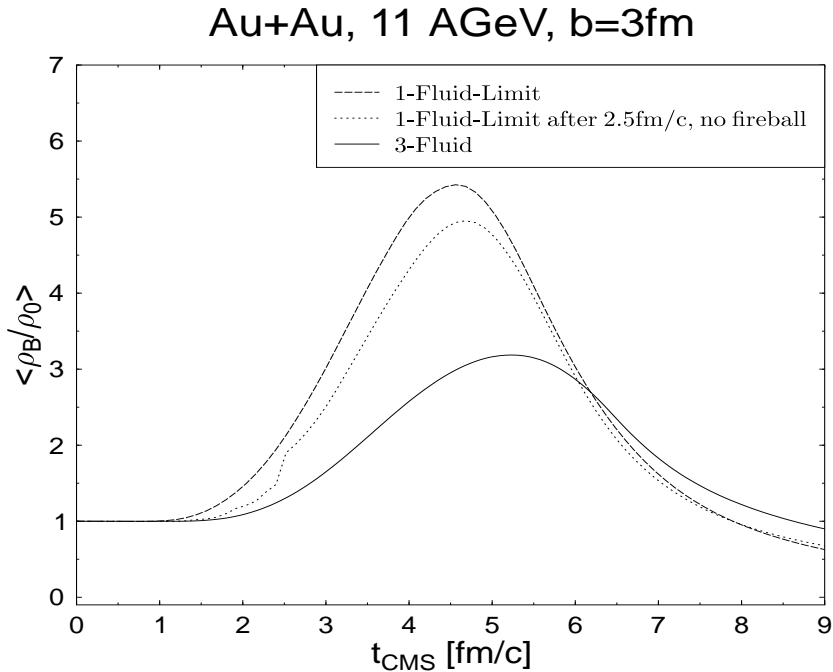,height=15cm,width=12cm}}}
\vspace*{-4cm}
\caption{Average baryon density of the projectile.}
\label{denspic}
\vspace*{.5cm}
\end{figure}

One-fluid calculations reach a phase transition to QGP
already at AGS energies or even below because the assumption
of instantaneous local equilibriation leads to maximal energy
deposition in the central region.
In the three-fluid model the finite stopping length 
of nuclear matter reduces the compression as can be seen in Fig.\
\ref{denspic}. In Fig.\ \ref{denspic} the evolution of the
compression of the projectile {\em only}, is compared to the 
one-fluid limit.
The maximum in the one-fluid limit is twice the compression in
the three-fluid model and is reached earlier.
The curve in between shows a calculation in which the creation
of the fireball is omitted by setting $f^\nu_{loss} = 0$, so that no
energy transfer to the third fluid is possible.
This also yields a higher compression than in the full three-fluid
calculation. At $t_{CMS}=2.5$ fm/c unification is enforced as in the
one-fluid limit. This results in a jump of the curve towards
the one-fluid limit calculation but does not reach the full height
since some of the impact energy is already deposited.

\begin{figure}[hbp]
\centerline{\hbox{
\psfig{figure=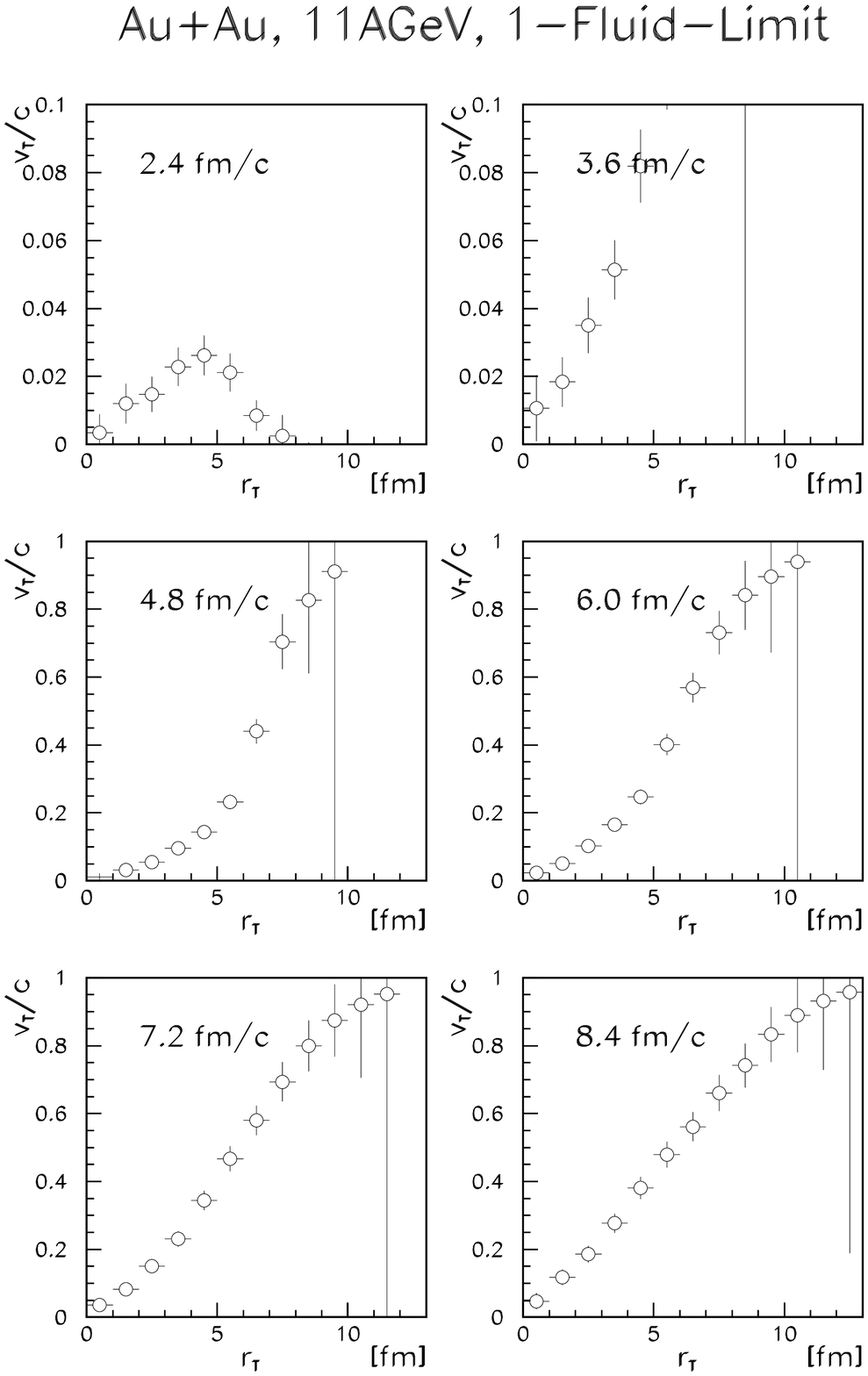,height=20cm,width=13cm}}}
\vspace*{-1.5cm}
\caption{The evolution of the transverse velocity profile 
in the one-fluid limit (b=$0~$fm).}
\label{betat1flimit}
\vspace*{.5cm}
\end{figure}

\begin{figure}[hbp]
\centerline{\hbox{
\psfig{figure=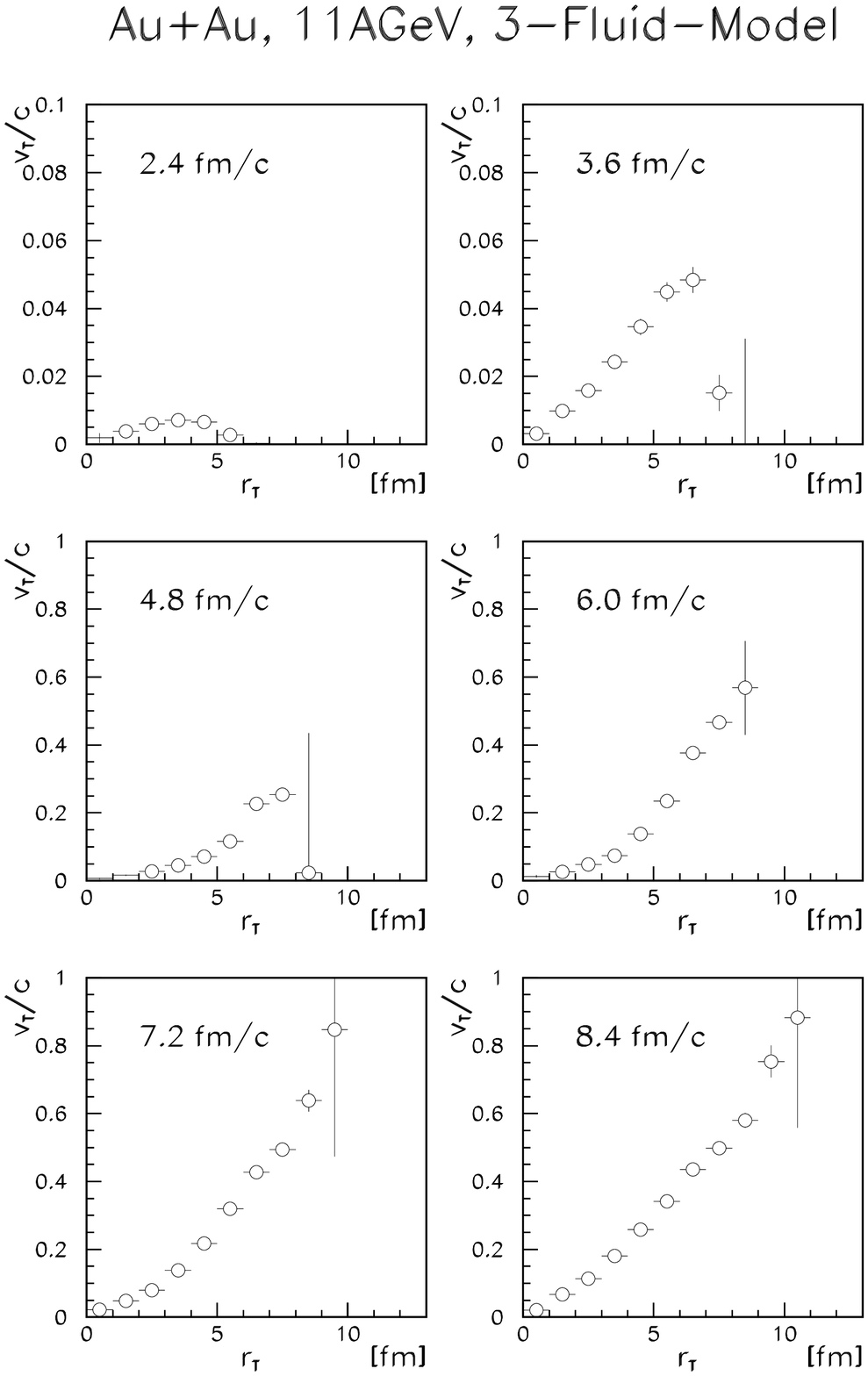,height=20cm,width=13cm}}}
\vspace*{-1.5cm}
\caption{The evolution of the transverse velocity profile 
in the three-fluid model (b=$0~$fm).}
\label{betat3fluid}

\vspace*{.5cm}
\end{figure}

As pointed out in the introduction, for a first order phase transition 
a longer lifetime or slower (transverse) expansion of the system is predicted
by the one-fluid model, which is due to the softened EoS.
A similar behaviour can be achieved by taking non-equilibrium effects
into account \cite{sorge}. They also soften the EoS since fewer
energy is deposited that could be converted into radial flow. Furthermore,
as long as the colliding matter is not yet thermalized,
only the partial and not the full equilibrium pressure is driving
the transverse expansion \cite{3fluid,gsi96}.
Figs.\ \ref{betat1flimit} and \ref{betat3fluid}
clearly show this effect. In the one-fluid limit the tranverse 
velocity profile reaches speed of light already at $4.8$~fm/c,
i.e.\ it accelerates faster into the radial direction 
than in the three-fluid model.
This results in a faster expansion, so that at later times
the one-fluid limit profiles reach out far beyond the $r_t=10.0$~fm.

\begin{figure}[hbp]
\centerline{\hbox{
\psfig{figure=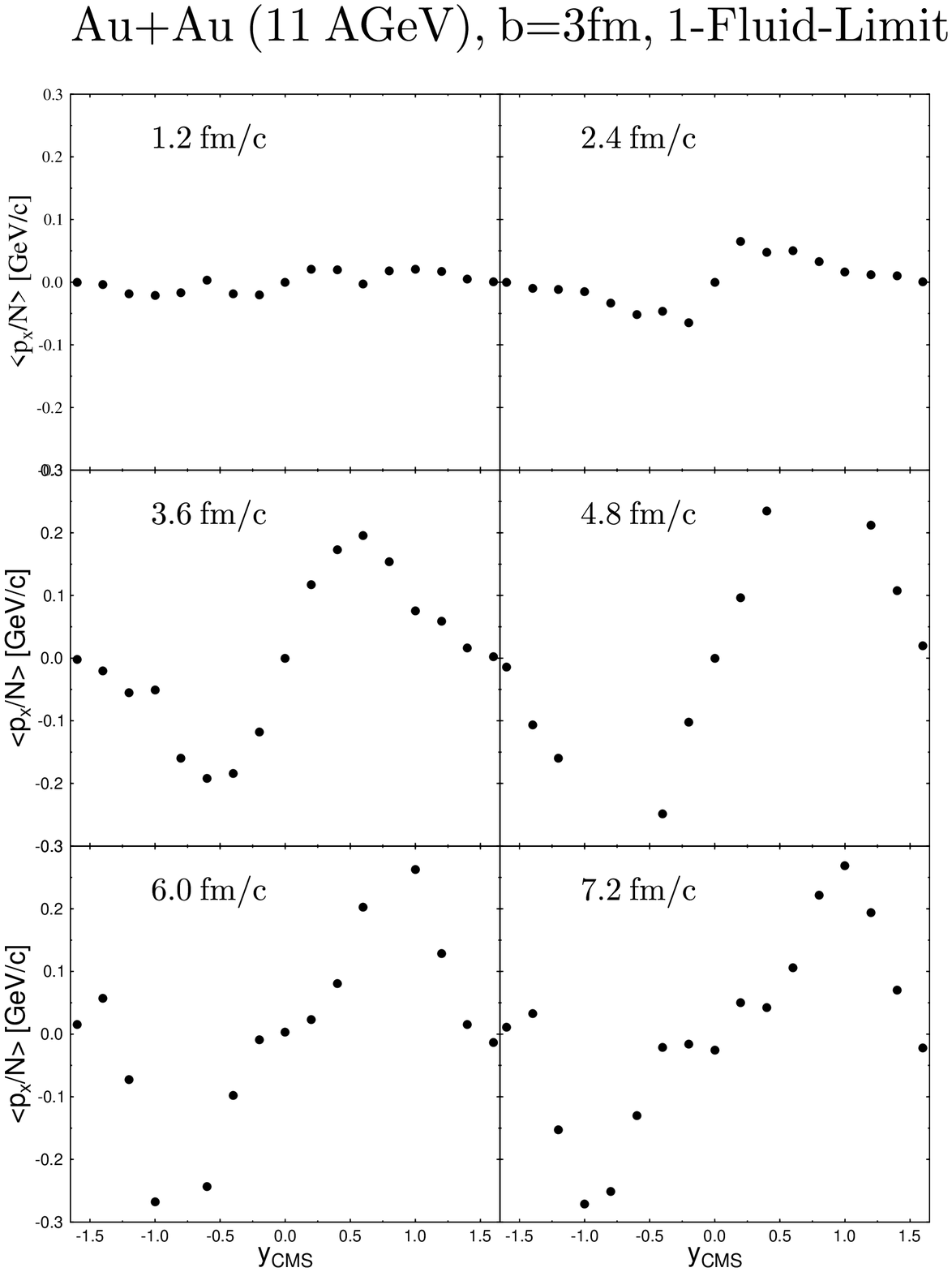,height=20cm,width=13cm}}}
\vspace*{-2cm}
\caption{The evolution of $\langle p_x/N\rangle$ in the one-fluid limit.}
\label{pxy1fluidpic}
\vspace*{.5cm}
\end{figure}

\begin{figure}[hbp]
\centerline{\hbox{
\psfig{figure=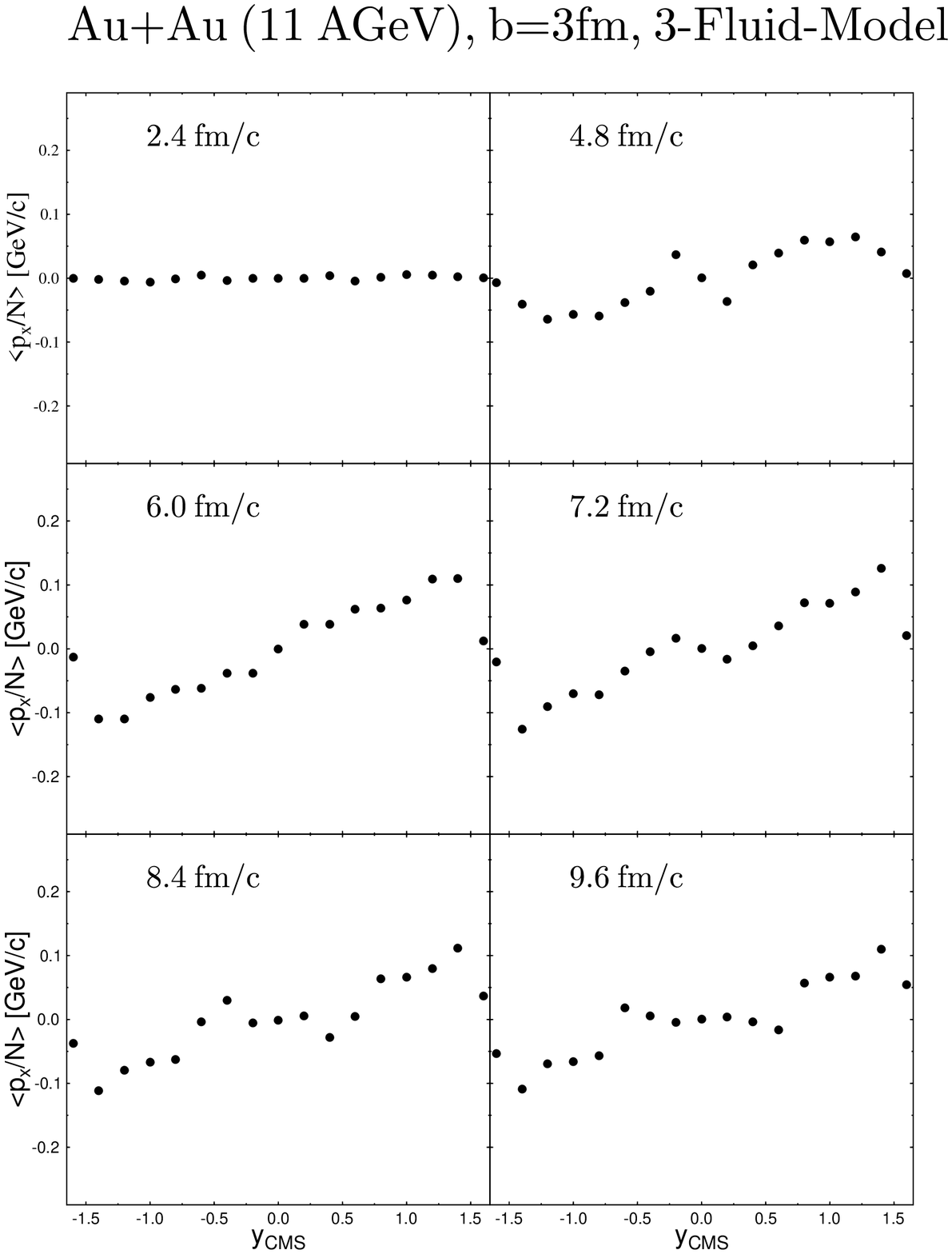,height=20cm,width=13cm}}}
\vspace*{-2cm}
\caption{The evolution of $\langle p_x/N\rangle$ in the three-fluid model.}
\label{pxy3fluidpic}
\vspace*{.5cm}
\end{figure}

The situation is similar when considering the in-plane directed flow.
The pressure build up in the central region can only bounce the
``spectator caps'' as long as they are passing this zone. 
In one-fluid hydrodynamics the directed flow is most sensitive to the
EoS of the {\em equilibrated} matter in the hot and dense central region.
Therefore, one-fluid calculations predict a significant minimum in the 
excitation function of the directed flow in case of a (first order)
phase transition to QGP.
However, the pressure during this stage can also considerably be lowered
by non-equilibrium effects \cite{sorge} for the same reasons as given in the
above discussion for the slower radial expansion.
Fig.\ \ref{pxy1fluidpic} shows the rapidity dependence of the
mean in-plane momentum per nucleon. It exhibits flow of up to
$\langle p_x/N\rangle(y) \approx  300~$MeV/c. 
In contrast, the flow in the three-fluid model
(Fig.\ \ref{pxy3fluidpic}) does not exceed $120~$MeV/c. 
Extracting the mean directed flow $p_x^{dir}$,
which is a weighted mean of the distributions depicted in Figs.\
\ref{pxy1fluidpic}, \ref{pxy3fluidpic}, the difference between
equilibrium (one-fluid limit) and non-equilibrium effects (three-fluid model)
with the crude EoS, eq.\ (\ref{EoS}), is of the same order of magnitude
as the difference between a one-fluid calculation using an EoS 
with or without a first order phase transition \cite{Yaris}.


\section{Summary and Outlook}
In this paper we presented a three-fluid hydrodynamical model which
allows to account for non-equilibrium effects between target, projectile,
and produced particles during the early stage of the collision.
We discussed that due to the nonvanishing thermalization time scale,
this model yields a
\begin{enumerate}
\item lower transverse pressure,
\item less baryonic compression,
\item and a different transverse velocity distribution
\end{enumerate}
of the nucleons {\em at early times} as compared to the one-fluid hydrodynamical
model (which assumes {\em instantaneous} local thermalization between
projectile, target, and produced particles). As a consequence, the
directed nucleon flow and the lifetime of the hot and dense central region
differ considerably in the three-fluid model as compared to the one-fluid model.
These results suggest that the predictions of the one-fluid model
may have to be modified by taking non-equilibrium effects into account,
if one assumes that other mechanisms, increasing the equilibration rate,
can be neglected.

\begin{figure}[hbp]
\centerline{\hbox{
\psfig{figure=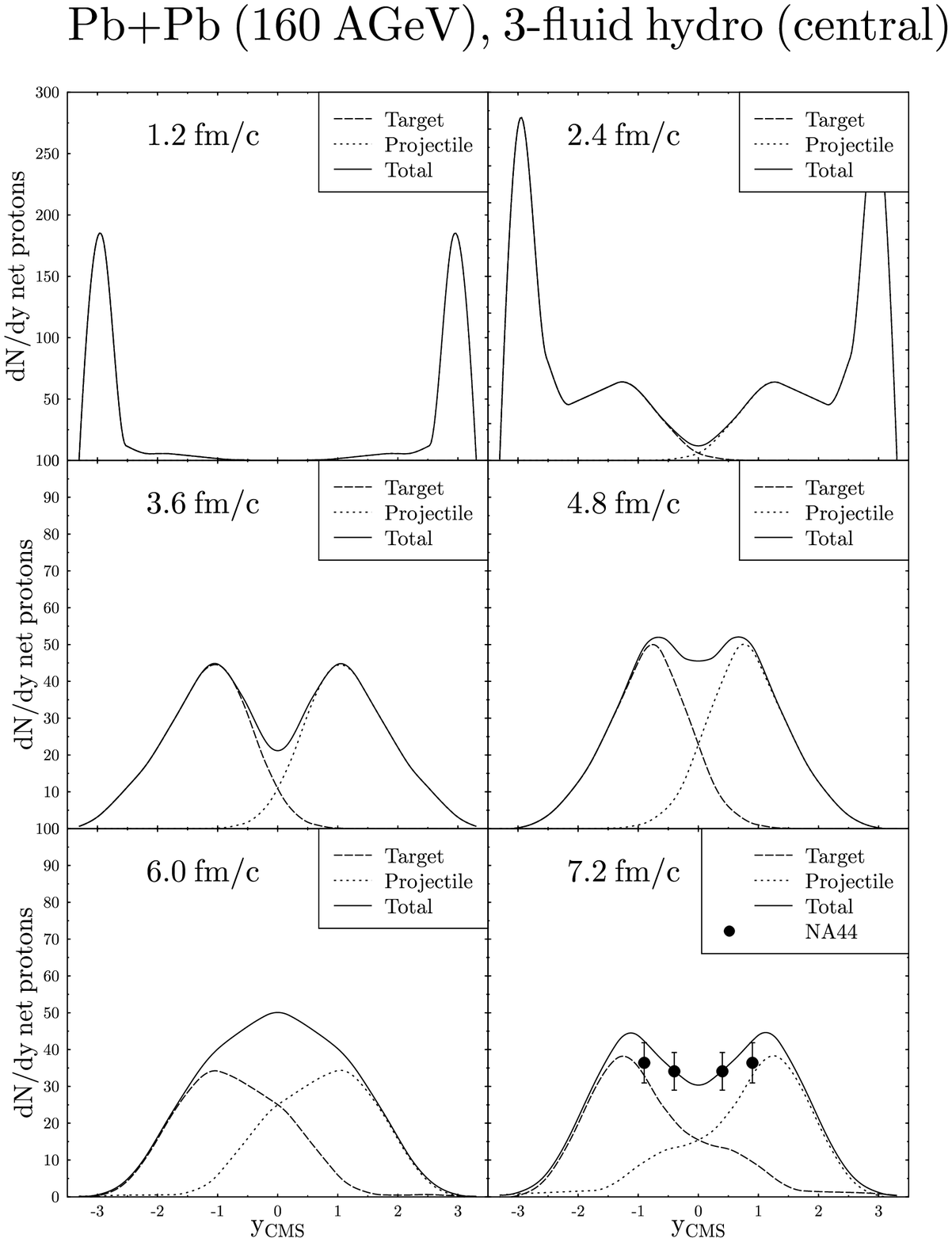,height=20cm,width=13cm}}}
\vspace*{-2cm}
\caption{The evolution of the proton dN/dY (obtained by scaling the 
net baryon multiplicity by $Z/A$ at all rapidities) 
in the three-fluid model}
\label{dndy160gev}

\vspace*{.5cm}
\end{figure}

In the future 
an excitation function of 
the directed flow will be calculated applying
a more refined EoS than considered here, in particular including a phase 
transition. 
The baryon dynamics (espescially flow) will also be studied at higher
beam energies (the rapidity and transverse momentum distributions
of pions in this model were already studied in \cite{Dum95a}).
The evolution of the (thermally smeared) rapidity distribution of protons 
in a Pb(160GeV)Pb collision is shown in Fig.\ \ref{dndy160gev}. 
Here, the non-equilibrium situation in the beginning of the reaction
becomes clear. The nucleonic fluids are not immediately stopped
at midrapidity but decellerate gradually. The comparison with
the (preliminary) NA44 data \cite{na44} supports that our
source terms yield sufficient stopping, even at such high energies.

\end{document}